\documentclass[a4paper,11pt]{article}
\pdfoutput=1 

\usepackage{jcappub} 
\usepackage[export]{adjustbox}
\usepackage[T1]{fontenc} 
\usepackage{amsmath}
\usepackage{tensor}
\usepackage{wasysym}
\usepackage{verbatim}
\usepackage{color}
\usepackage{mathtools,slashed}
\usepackage{subcaption}

\newcommand{\Beq}{\begin{equation}\begin{aligned}}
\newcommand{\Eeq}{\end{aligned}\end{equation}}
\usepackage{braket}
\usepackage[utf8]{inputenc} 
\usepackage[T1]{fontenc}

\title{Effects of Gravitational Chern-Simons during Axion-SU(2) Inflation}

\author[a,1]{Leila Mirzagholi,\note{Corresponding author.}}
\author[a,b]{Eiichiro Komatsu,}
\author[a]{Kaloian D. Lozanov}
\author[a,c]{and Yuki Watanabe}

\affiliation[a]{Max Planck Institute for Astrophysics, Karl-Schwarzschild-Str.1, 85741 Garching, Germany}
\affiliation[b]{Kavli Institute for the Physics and Mathematics of the Universe (Kavli IPMU, WPI), UTIAS, The University of Tokyo, Chiba 277-8583, Japan}
\affiliation[c]{Department of Physics, National Institute of Technology, Gunma College, Gunma 371-8530, Japan}

\emailAdd{lmirz@mpa-garching.mpg.de}
\emailAdd{komatsu@mpa-garching.mpg.de}
\emailAdd{klozanov@mpa-garching.mpg.de}
\emailAdd{yuki.watanabe@gunma-ct.ac.jp}

\abstract{In this paper we examine the viability of inflation models with a spectator axion field coupled to both gravitational and SU(2) gauge fields via Chern-Simons couplings. Requiring phenomenological success of the axion-SU(2) sector constrains the coupling strength of the gravitational Chern-Simons term. We find that the impact of this term on the production and propagation of gravitational waves can be as large as fifty percent enhancement for the helicity that is not sourced by the gauge field, if the cut-off scale is as low as $\Lambda = 20H$. The effect becomes smaller for a larger value of $\Lambda$, while the impact on the helicity sourced 
by the gauge field is negligible regardless of $\Lambda$.}

\begin{document}
\maketitle
\flushbottom
\section{Introduction}

Coupled axion and SU(2) gauge fields during inflation \cite{Maleknejad:2011jw,Maleknejad:2011sq,Adshead:2012kp,Adshead:2013nka} provide a rich phenomenology that is not shared by canonical single scalar field inflation \cite{Guth:1980zm,Sato:1980yn,Linde:1981mu,Albrecht:1982wi} models (see \cite{Maleknejad:2012fw} for a review), e.g., they generate a stochastic background of \textit{chiral} gravitational waves \cite{Adshead:2013qp, Maleknejad:2012fw, Dimastrogiovanni:2012ew, Maleknejad:2016qjz} which are non-Gaussian \cite{Agrawal:2017awz,Agrawal:2018mrg, Fujita:2018vmv, Dimastrogiovanni:2018xnn}. It has also been shown \cite{Alexander:2004us, Maleknejad:2016dci, Adshead:2017znw, Caldwell:2017chz, Papageorgiou:2017yup, Adshead:2018doq, Noorbala:2012fh, Maleknejad:2014wsa} that generation of chiral gravitational waves leads to a non-zero parity-violating gravitational anomaly, $R\tilde{R}$, which, in turn, violates the lepton number conservation and generates the baryon asymmetry of the universe.

The upcoming LiteBIRD \cite{Matsumura_2014, Hazumi:2019lys} and CMB Stage-4 experiments \cite{Abazajian:2016yjj} will provide further constraints on the axion-gauge fields \cite{Thorne:2017jft, Shandera:2019ufi}. However, before the predictions of these models are taken seriously, it is important to check if these models are viable both phenomenologically and theoretically.  

From the phenomenological point of view, it is important to see if couplings of SU(2) gauge fields to other fields lead to particle production and their backreaction on the axion-gauge field backgrounds do not spoil the model setup. In \cite{Lozanov:2018kpk}, a charged scalar field is coupled to the SU(2) gauge field, and production and backreaction of pairs of charged particles is studied in de Sitter spacetime. In \cite{Maleknejad:2018nxz}, the backreaction of the extra spin-2 field in this setup is analytically studied for all the inflationary models involving the SU(2) gauge field. In \cite{Mirzagholi:2019jeb, Maleknejad:2019hdr} a pair of massive Dirac fermions are coupled to the SU(2) gauge field. The coupling to a massless fermion is studied in \cite{Domcke:2018gfr}. In all these cases there exists a parameter space in which the backreaction of the particles on the SU(2) background is negligible. The nonlinear impact on the scalar perturbations of the chromo-natural inflation and the spectator sector during inflation has been studied in \cite{Papageorgiou:2018rfx, Papageorgiou:2019ecb}. Anisotropic initial conditions are discussed in \cite{Wolfson:2020fqz}. 

From the theoretical point of view, it is important to consider all different classes of parity-violating terms that arise from the same physics. The existence of both $F\tilde{F}$ and $R\tilde{R}$ is the result of one phenomenon. As an example, if coupling to a massive degree of freedom, such as  axially coupled heavy fermions, is considered, then radiative fermion loops generate not only $F\tilde{F}$, but also $R\tilde{R}$ \cite{Lue:1998mq}. String theory predicts the existence of axions that couple to both terms simultaneously \cite{Witten:1984dg, Choi:1997an, Choi:1999zy}. Hence, these two parity-violating terms arise at the same time and should be effectively considered on the same level in the theory. Moreover, given that $R\tilde{R}$ can introduce a ghost instability, it is important to check the stability of these models up to their cut-off scales. 

In this paper we consider the axion-SU(2) gauge field spectator sector \cite{Dimastrogiovanni:2016fuu} together with the gravitational Chern-Simons term coupled to the axion field. In this setup we have the parity-violating terms on both sides of the equation of motion for the tensor metric perturbations; the left-hand side due to the gravitational Chern-Simons term and the right-hand side due to the axion-SU(2) sector. There are many studies focused on the cosmological signatures of the gravitational Chern-Simons term \cite{Lue:1998mq, Jackiw:2003pm, Alexander:2004wk, Lyth:2005jf, Saito:2007kt, Satoh:2007gn, Fischler:2007tj, Alexander:2009tp, Dyda:2012rj, Bartolo:2017szm, Bartolo:2018elp, Kamada:2019ewe, Qiao:2019hkz}. In this paper, we extend these studies to the axion-SU(2) models. In \cite{Basilakos:2019mpe, Basilakos:2019wxu, Basilakos:2019acj, Basilakos:2020qmu}, the authors have considered both $F\tilde{F}$ and $R\tilde{R}$ originating from string theory. The non-abelian gauge field in their consideration does not share the same vacuum expectation value as in our case, hence it does not source gravitational waves linearly. 

This paper is organized as follows. In section \ref{sec:setup} we present the model and analyse the background evolution. In section \ref{sec:tp} we present the second-order Lagrangian for tensor perturbations and compare the gravitational waves with and without the gravitational Chern-Simons term. We discuss the stability and the cut-off scale of the model in section \ref{sec:stab}. We conclude in section \ref{sec:dis}.

\section{Model Setup} 
\label{sec:setup}
\subsection{Action}
We consider the following action
\Beq 
\label{action}
S = S_{EH} + S_{\varphi} + S_{SPEC} + S_{GCS} \,,
\Eeq
where $S_{EH}$ is the Einstein-Hilbert action and $S_{\varphi}$ is the inflaton sector action given by 
\Beq
S_{EH} = \int d^4x \sqrt{-g} \frac{M_{pl}^2}{2}R\,,
\Eeq
\Beq
S_{\varphi} = \int d^4x \sqrt{-g}\Big( - \frac{1}{2}(\partial \varphi)^2 - V(\varphi)\Big)\,.
\Eeq

The spectator sector action $S_{SPEC}$ contains axion and SU(2) gauge fields, where $\chi$ is an axion field with potential $U(\chi)$ and a decay constant $f$:
\Beq
\label{SPEC}
\begin{split}
S_{SPEC} = \int d^4x \sqrt{-g}\Big(- \frac{1}{2}(\partial \chi)^2 - U(\chi) - \frac{1}{4} F^a_{\mu \nu}F^{a \mu \nu}+ \frac{\lambda_1 \chi}{4f} F^a_{\mu \nu}\tilde{F}^{a \mu \nu} \Big)\,,
\end{split}
\Eeq
where
\Beq
F^a_{\mu\nu}=\partial_{\mu}A^a_{\nu}-\partial_{\nu}A^a_{\mu}-g_{\!A}\epsilon^{abc}A^b_{\mu}A^c_{\nu}\,,
\Eeq
is the field strength tensor of the SU(2) gauge fields, with $g_A$ being the self-coupling constant and $\epsilon^{abc}$ the three dimensional anti-symmetric symbol. 

The last term in $S_{SPEC}$ is the Chern-Simons interaction, where $\lambda_1 / f$ parametrizes its coupling strength and $\tilde{F}^{a}{}^{\mu\nu} \equiv \varepsilon^{\mu \nu \alpha \beta}F^a_{\alpha \beta}/2$ is the dual of $F^a_{\mu\nu}$. $\varepsilon^{\mu \nu \alpha \beta}$ is defined as $\varepsilon^{\mu \nu \alpha \beta} \equiv \epsilon^{\mu \nu \alpha \beta}/\sqrt{-g}$, where $\epsilon^{\mu \nu \alpha \beta}$ is the totally anti-symmetric symbol with $\epsilon^{0123}=1$. The $S_{SPEC}$ is invariant under the local SU(2) transformation. $F\tilde{F}$ is a total derivative and for $\chi = const$, it reduces to a surface term. Hence, we can write $F\tilde{F}$ as
\Beq
F\tilde{F} = \nabla_{\mu}C^{\mu}\,,
\Eeq
with
\Beq
C^{\mu} = 2 \varepsilon^{\mu \nu \alpha \beta}  \Big( A^a_{\nu}\partial_{\alpha} A^a_{\beta} - \frac{1}{3} \epsilon^{abc} A^a_{\nu} A^b{_\alpha} A^c_{\beta}\Big)\,. 
\Eeq

The last term in \eqref{action} is the gravitational Chern-Simons term coupled to the axion, with coupling strength of $\lambda_2 / f$:
\Beq
\label{GCS}
S_{GCS} =  \int d^4x \sqrt{-g}  \frac{\lambda_2\chi}{4f} R\tilde{R}\,, 
\Eeq
where
\Beq 
R\tilde{R} =\tensor{R}{^\beta _\alpha ^{\gamma \delta}}\tensor{\tilde{R}}{^\alpha _{\beta \gamma \delta}}\,,
\Eeq
$\tensor{R}{^\beta _\alpha ^{\gamma \delta}}$ is the Riemann tensor and the dual of the Riemann tensor is
\Beq
\tensor{\tilde{R}}{^\alpha _{\beta \gamma \delta}} = (1/ 2)\varepsilon_{\sigma \tau\gamma \delta}\tensor{R}{^\alpha _\beta ^{\sigma \tau}}\,.
\Eeq
We can also write $R\tilde{R}$ as
\Beq
R\tilde{R} = 2 \nabla_{\mu}K^{\mu}\,,
\Eeq
with
\Beq
K_{\mu} = 2 \varepsilon^{\mu \alpha \beta \gamma} \Big( \frac{1}{2}\Gamma^{\sigma}_{\alpha \tau} \partial_{\beta} \Gamma^{\tau}_{\gamma \sigma} + \frac{1}{3} \Gamma^{\sigma}_{\alpha \tau} \Gamma^{\tau}_{\beta \eta} \Gamma^{\eta}_{\gamma \sigma}\Big)\,.
\Eeq
The gravitational Chern-Simons term is also a total derivative. For $\chi = const$, it reduces to a surface term. 
\subsection{Background Evolution} 
\label{sec:bg}
In this section, we describe the evolution of the background. As $R\tilde{R}$ vanishes in a Friedmann-Lema\^itre-Robertson-Walker (FLRW) background, there is no contribution to the background.

The vacuum expectation value of the gauge field is given by \cite{Maleknejad:2011jw, Maleknejad:2011sq}
\Beq
A^a_0 = 0, \quad A^a_i = \delta_i^a a(t) Q(t)\,.
\Eeq
The 00-component of the Einstein equations is \cite{Dimastrogiovanni:2016fuu} 
\Beq
3M_{pl}^2 H^2 = \frac{\dot{\varphi}^2}{2} + V(\varphi) + \frac{\dot{\chi}^2}{2} + U(\chi) + \frac{3}{2}(\dot{Q}+ HQ)^2 + \frac{3}{2}g_{\!A}^2 Q^4\,. \Eeq
The equations of motion for the axion and gauge fields are given by \cite{Adshead:2012kp, Dimastrogiovanni:2016fuu}
\Beq
\label{eq:axion}
\ddot{\chi} + 3H\dot{\chi} + \partial_{\chi}U = - 3\frac{g_{\!A} \lambda_1}{f} Q^2 (\dot{Q} + HQ)\,,
\Eeq
\Beq
\label{eq:gauge}
\ddot{Q} + 3H\dot{Q} + (\dot{H} + 2H^2) Q + 2g_{\!A}^2Q^3 = \frac{g_{\!A} \lambda_1}{f} \dot{\chi} Q^2\,,
\Eeq
where dots show derivatives with respect to the cosmic time $t$ and $H \equiv \dot{a}/a$ is the Hubble expansion rate. 
It is useful to define the slow-roll parameters $\epsilon_H \equiv - \dot{H}/H^2$ and write \cite{Dimastrogiovanni:2016fuu} 
\Beq
\epsilon_H = \epsilon_{\varphi} + \epsilon_{\chi} + \epsilon_B + \epsilon_E\,,
\Eeq
where
\Beq
\epsilon_{\varphi} \equiv \frac{\dot{\varphi}^2}{2H^2M_{pl}^2}, \quad \epsilon_{\chi} \equiv \frac{\dot{\chi}^2}{2H^2M_{pl}^2}, \quad \epsilon_{B} \equiv \frac{g_{\!A}^2Q^4}{H^2M_{pl}^2}, \quad \epsilon_{E} \equiv \frac{(\dot{Q}+HQ)^2}{H^2M_{pl}^2}\,,
\Eeq
are all much smaller than unity.

Also we define the following dimensionless parameters
\Beq
\label{ndparameters}
m_{Q} \equiv \frac{g_{\!A}Q}{H}, \quad \xi_1 \equiv \frac{\lambda_1 \dot{\chi}}{2fH}, \quad \xi_2 \equiv \frac{1}{2} \frac{\lambda_2 \dot{\chi}}{2f H}\bigg(\frac{H}{M_{pl}}\bigg)^2\,.
\Eeq
The fourth term in the left hand side of \eqref{eq:gauge} becomes $2m_Q^2H^2Q$; thus $m_Q$ can be regarded as the mass of $Q$ (divided by $H$). 

In the slow-roll approximation, the following relation holds between $m_{Q}$ and $\xi_1$\cite{Dimastrogiovanni:2016fuu}
\Beq
\xi_1  \simeq m_{Q} + \frac{1}{m_Q}\,.
\Eeq

To prevent instabilities of the scalar perturbations (lower bound) and strong backreaction on the gauge background (upper bound), we consider $\sqrt{2} < m_{Q} \lesssim 4$ \cite{Dimastrogiovanni:2012ew, Agrawal:2018mrg, Dimastrogiovanni:2016fuu, Maleknejad:2018nxz}. This implies 
\Beq
\xi_2 \simeq \frac{\lambda_2}{2\lambda_1}\bigg(\frac{H}{M_{pl}}\bigg)^2 \big(m_{Q} + \frac{1}{m_Q}\big) \lesssim 10^{-9} \bigg(\frac{\lambda_2}{\lambda_1}\bigg)\,,
\Eeq
where we have used $\big(H/M_{pl}\big)^2 \lesssim 10^{-9}$. Thus, a sizeable $\xi_2$, e.g., $\xi_2 \simeq 10^{-2}$, requires large $\lambda_2 / \lambda_1$, e.g., $\big(\lambda_2 / \lambda_1\big)= 10^7$. A large hierarchy between $\lambda_2$ and $\lambda_1$ i.e., $\big(\lambda_2 / \lambda_1\big) \gg 1$, is in principle allowed, since all the degrees of freedom are coupled to the gravitational Chern-Simons term, but only the charged ones are coupled to the SU(2) Chern-Simons term.  Specifically, $\lambda_2  \propto (f/\Lambda) N$ where $N$ is the number of integrated out degrees of freedom and $\Lambda$ is the cut-off scale of the effective field theory, e.g., the mass of fermions in the loops. Note that this holds assuming that we are integrating out the massive fermions as an example at the same energy scale to get $F\tilde{F}$ and $R\tilde{R}$ simultaneously. 

\section{Tensor Perturbations}
\label{sec:tp}
Now, we consider tensor perturbations during inflation at linear level. The tensor perturbations are amplified due to the tachyonic instability, whereas the scalar and vector perturbations are not amplified for $m_Q > \sqrt{2}$ \cite{Dimastrogiovanni:2016fuu, Maleknejad:2012fw, Adshead:2013nka, Dimastrogiovanni:2012ew} in the spectator axion-SU(2) sector. 

We have four tensor degrees of freedom: two metric tensor degrees of freedom that represent the gravitational waves and two additional tensor degrees of freedom associated with the SU(2) gauge field. We consider a perturbed FLRW metric as follows
\Beq
ds^2 = a^2(\tau) (-d\tau ^2+ (\delta_{ij} + \tilde{h}_{ij}) dy^i dy^j)\,,
\Eeq
where $\tau \simeq - 1 / aH$ is the conformal time and $\tilde{h}_{ij}$ is a transverse and traceless tensor, i.e. $\partial^i\tilde{h}_{ij} = \tilde{h}^i_i = 0$. 
We define the Fourier transformed right and left-handed circular polarization states as
\Beq
\tilde{h}_{ij}(\tau ,y) = \int \sum_{A = L,R} \frac{d^3k}{(2\pi)^{3/2}} e^{A}_{ij}(k) \tilde{h}_{A}(\tau , k)e^{ik.y} \,,
\Eeq
where $e^{A}_{ij}$ is the polarization state tensor for the right $(A=R)$ and left-handed $(A=L)$ circular polarization states and satisfies the relation
\Beq
i k_a \epsilon^{ab}{}_ce^{R}_{db} = k e^{R}_{cd}, \quad
i k_a \epsilon^{ab}{}_ce^{L}_{db} = -k e^{L}_{cd} \,,
\Eeq
where $ \epsilon^{ab}{}_c$ is the three dimensional anti-symmetric symbol. For simplicity, we assume that the gravitational waves are propagating along the $z$ spatial direction
\Beq
ds^2 = a^2(\tau) \Big[-d\tau ^2+ (1 + \tilde{h}_+(\tau , z)) dx^2 + (1-\tilde{h}_+(\tau, z)) dy^2 + 2\tilde{h}_{\times}(\tau , z)dxdy +dz^2 \Big]\,.
\Eeq
We write the tensor perturbations of the gauge field as $\delta A^a_i = a\tilde{t}^a_i$, where $\tilde{t}^a_i$ is chosen to be transverse and traceless, i.e. $\partial_i\tilde{t}_{i}^{a} = \tilde{t}^{ai}_i = 0$.   We write the gauge tensor perturbations as $\delta A^1_i = a(\tilde{t}_+, \tilde{t}_{\times}, 0)$ and $\delta A^2_i = a(\tilde{t}_{\times}, -\tilde{t}_+, 0)$. For our convenience we work with the canonically normalised tensor perturbations 
\Beq
h_{ij} \equiv a\frac{M_{pl}}{\sqrt{2}}\tilde{h}_{ij}, \quad t_i^a =\sqrt{2} a \tilde{t}^a_i\,. 
\Eeq
We define the left and the right helicities as \cite{Dimastrogiovanni:2016fuu}
\Beq
h_{L,R} \equiv \frac{1}{\sqrt{2}}( h_{+} \pm i h_{\times}), \quad t_{L,R} \equiv \frac{1}{\sqrt{2}}(t_{+} \pm it_{\times})\,.
\Eeq

Now we write the second order action for the tensor perturbations. 
\Beq
\label{S2R}
S = \frac{1}{2}\sum_{A = L, R} \int d\tau d^3k \Bigg\{ \Big(1 - \frac{s\lambda_2 \chi'}{4faM_{pl}^2} \frac{k}{a}\Big) \Big[h_A^{'\dagger}h_A' + (-k^2+2\mathcal{H}^2)h_A^{\dagger}h_A- 2\mathcal{H}\Re(h_A^{'\dagger}h_A )\Big] \\  
+ t_A^{\dagger}t_A  \Big[ s ak\left(2g_{\!A}Q + \frac{\lambda_1}{f}a \chi'\right) -k^2 - \frac{g_{\!A} a Q \lambda_1}{f}\chi'\Big]  + 2\Re\Big[h^{' \dagger}_A t_A -h_A^{\dagger} t'_A\Big]\left(\frac{Q'+\mathcal{H}Q}{M_{pl}}\right)  \\  + t_A^{'\dagger}t_A'+ 2\Re(h_At_A^{\dagger})\Big[-s ak \frac{2g_{\!A}Q^2}{M_{pl}} - \frac{\mathcal{H}(Q'+\mathcal{H}Q)}{M_{pl}} + \frac{g_{\!A} \lambda_1 Q^2 a}{fM_{pl}}\chi'\Big] \Bigg\} \,,
\Eeq
where $s = -1,1$ for the left- and right-handed helicities respectively. A prime denotes the derivative with respect to the conformal time $\tau$, and $\mathcal{H} \equiv a'/a$.\\

Using the parameters defined in \eqref{ndparameters} in the action, we find 

\Beq
\label{secondorderlag}
S=  \frac{1}{2}\sum_{A = L, R}  \int d\tau d^3k \Bigg\{ \Big(1 - s\frac{\xi_2}{\mathcal{H}}k\Big) \Big[h_A^{'\dagger}h_A'+ h_A^{\dagger}h_A( -k^2+2\mathcal{H}^2)-2\mathcal{H} \Re(h_A^{'\dagger}h_A )\Big]
 \\
 + \Big(2\Re(h^{' \dagger}_A t_A) -2\Re(h_A^{\dagger} t'_A)\Big) \mathcal{H}\sqrt{\epsilon_E}
+  t_A^{\dagger}t_A  \Big[ 2 s\mathcal{H}m_Q k  + 2s \mathcal{H}\xi_1 k - 2\mathcal{H}^2m_Q \xi_1 -k^2\Big] \\
+ t_A^{'\dagger}t_A' + 2\Re(h_At_A^{\dagger})\Big(-2s \mathcal{H}\sqrt{\epsilon_B} k  + 2 \mathcal{H}^2 \sqrt{\epsilon_B} \xi_1  - \mathcal{H}^2 \sqrt{\epsilon_E}\Big) \Bigg\}\,.
\Eeq 

\begin{figure}[t]
\begin{subfigure}{0.5\textwidth}
\includegraphics[width=0.99\linewidth, height=5.4cm]{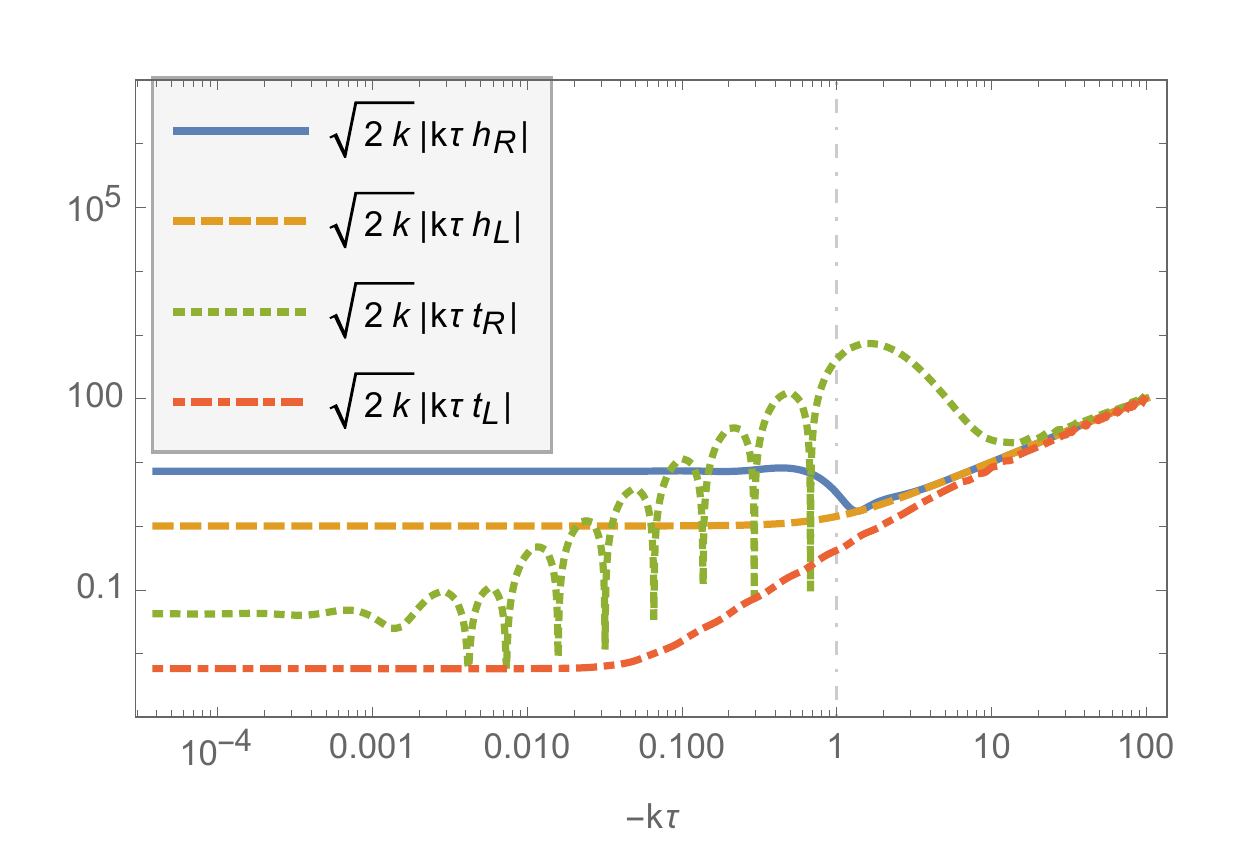} 
\end{subfigure}
\begin{subfigure}{0.5\textwidth}
\includegraphics[width=0.99\linewidth, height=5.4cm]{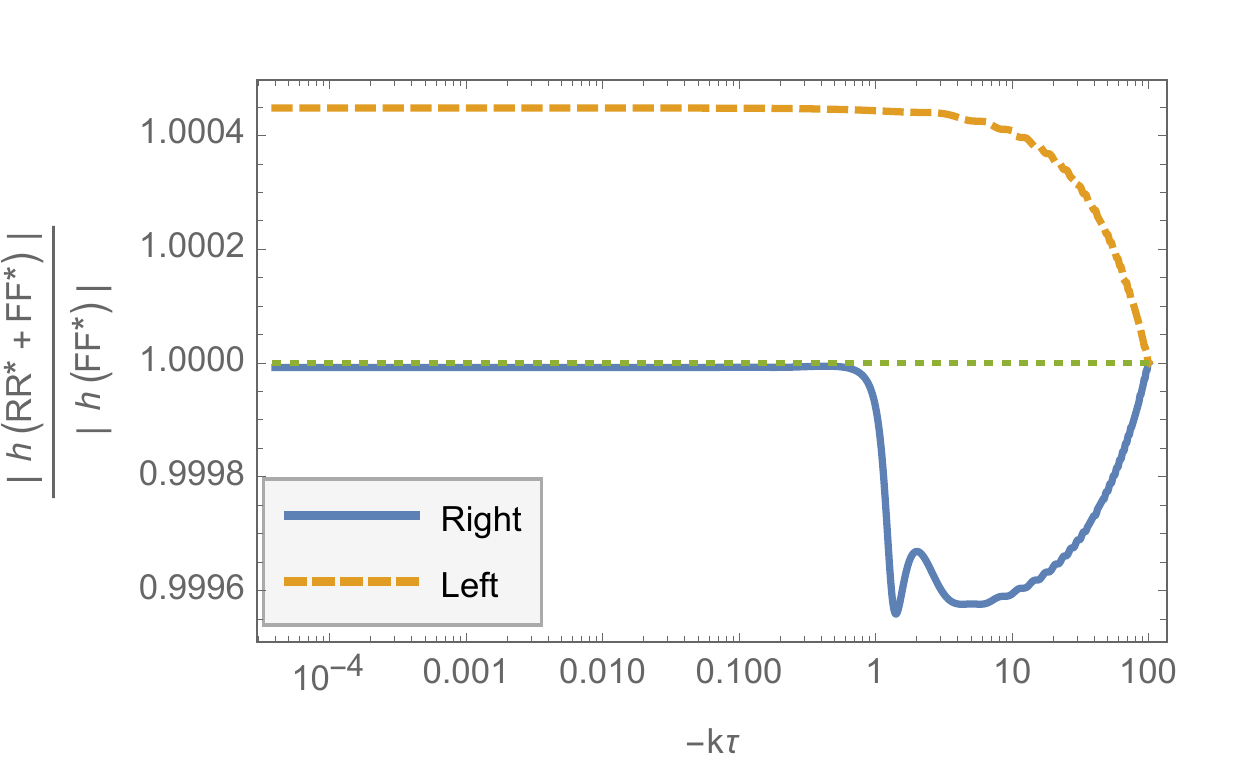}
\end{subfigure}
\caption{\textbf{(Left panel)} The sourced
linear gravitational wave $\sqrt{2k}|k\tau h_R|$ (solid blue), the left-handed gravitational waves  $\sqrt{2k}|k\tau h_L|$ (dashed orange) and the linear right- and left-handed gauge tensor mode functions $\sqrt{2k}|k\tau t_R|$ (dotted green) and $\sqrt{2k}|k\tau t_L|$ (dot-dashed red) are shown. The right-handed helicity gravitational wave grows around the horizon crossing and stays constant afterwards. \textbf{(Right panel)} The ratios of the right- (solid blue) and left-handed (dashed orange) helicity mode functions for $\xi_2 = 4.5\times10^{-6}$ with respect to those for $\xi_2 = 0$. The horizontal dotted line shows unity.}
\label{fig:i}
\end{figure}
\pagebreak
The equations of motion for the tensor modes up to leading order in slow-roll parameters are
\Beq
\label{eomtR}
\partial_x^2 t_A + \Big[1+\frac{2}{x^2} \big(m_Q \xi_1 - s x(m_Q + \xi_1)\big)\Big] t_A
= - \frac{2\sqrt{\epsilon_E}}{x} \partial_x h_A + \frac{2}{x^2} \Big[(m_Q  - s x)\sqrt{\epsilon_B} + \sqrt{\epsilon_E} \Big]h_A\,,\\
\Eeq
\Beq
\label{eomhR}
\Big(1 -  s \xi_2 x\Big) \Big[ \partial_x^2 h_A +(1 - \frac{2}{x^2} )h_A \Big] - 2 s \xi_2 \partial_x h_A
= \frac{2\sqrt{\epsilon_E}}{x} \partial_x t_A + \frac{2}{x^2} \Big[(m_Q  -s x)\sqrt{\epsilon_B} \Big]t_A\,,
\Eeq
where $x \equiv -k\tau$. 

Next we calculate the four tensor modes numerically.\footnote{For a discussion of the quantisation of the tensor modes and their vacuum state, see Appendix \ref{initial}.} Only the right-handed helicity mode of the tensor perturbations of the SU(2) gauge field $t_R$ is amplified unlike the gravitational waves. 
In the left panel of Figure \ref{fig:i} we show the amplification of $|t_R| = \sqrt{\braket{ t_R^{\dagger}t_R}}$ (green line) around the horizon crossing $(|k\tau| \sim 1)$, assuming $m_Q$, $\xi_1$, $\xi_2$, $H$, $\epsilon_B$ and $\epsilon_E$ are constant. In all the plots in this section we use the following parameters
\Beq
m_{Q} = 3, \quad  \epsilon_B = 3 \times 10^{-4}, \quad  \epsilon_E  = 3 \times 10^{-5}, \quad H=10^{13} \text{GeV}\,.
\Eeq

\subsection{Without $F\tilde{F}$}
\label{WFF}
\begin{figure}[ht]
\begin{subfigure}{0.5\textwidth}
\includegraphics[width=0.99\linewidth, height=5.4cm]{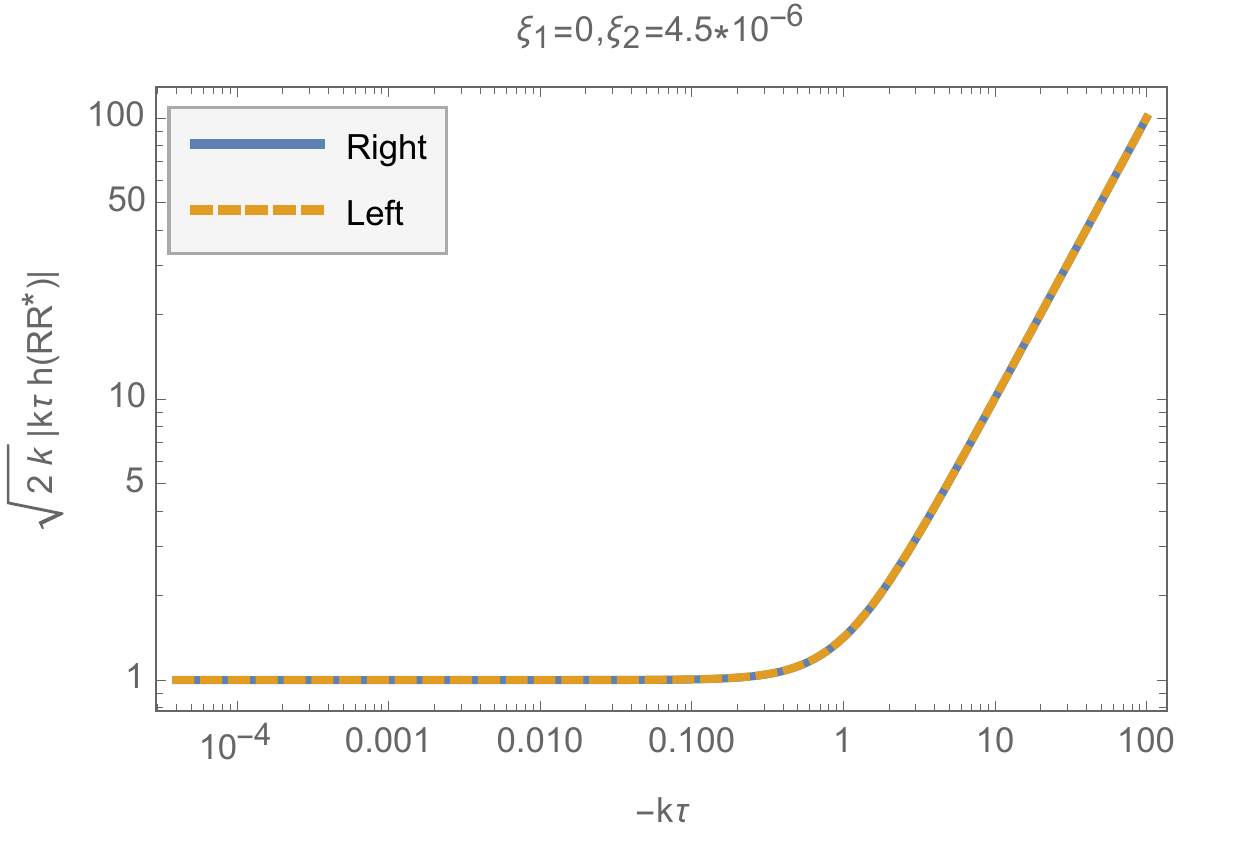} 
\end{subfigure}
\begin{subfigure}{0.5\textwidth}
\includegraphics[width=0.99\linewidth, height=5.4cm]{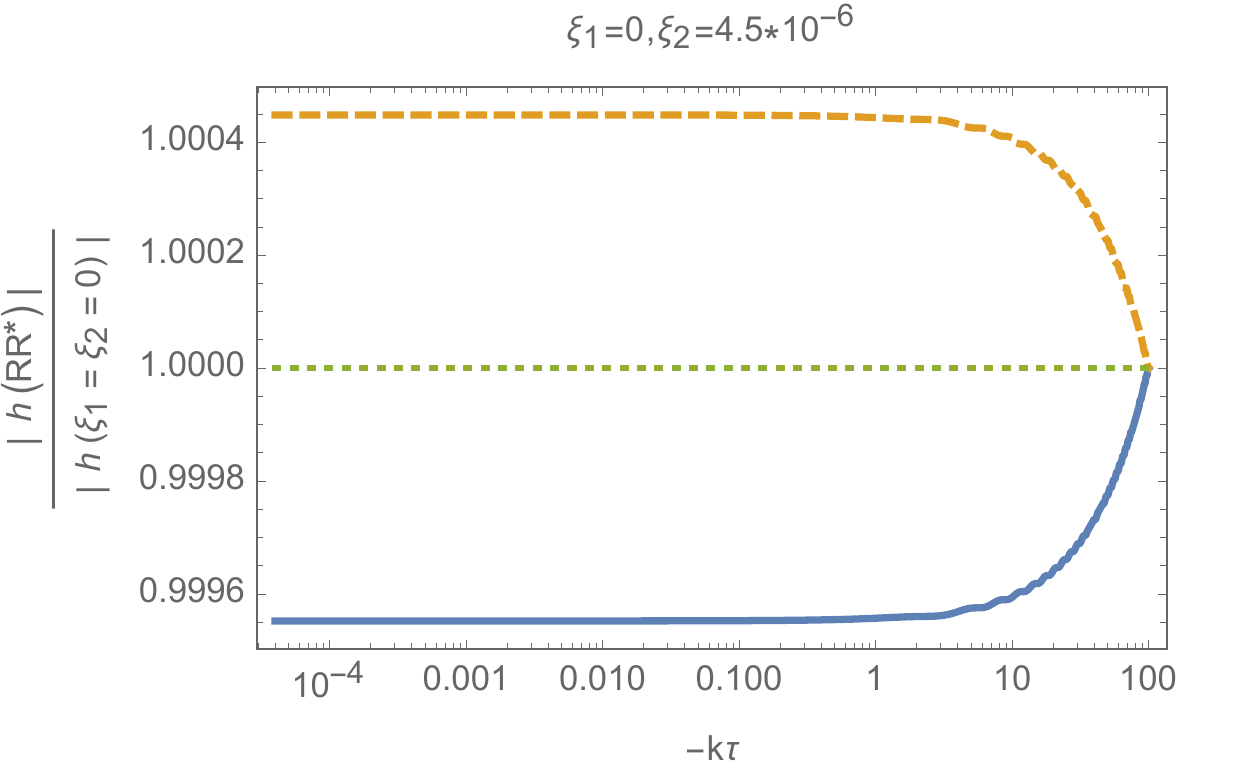}
\end{subfigure}
\begin{subfigure}{0.5\textwidth}
\includegraphics[width=0.99\linewidth, height=5.4cm]{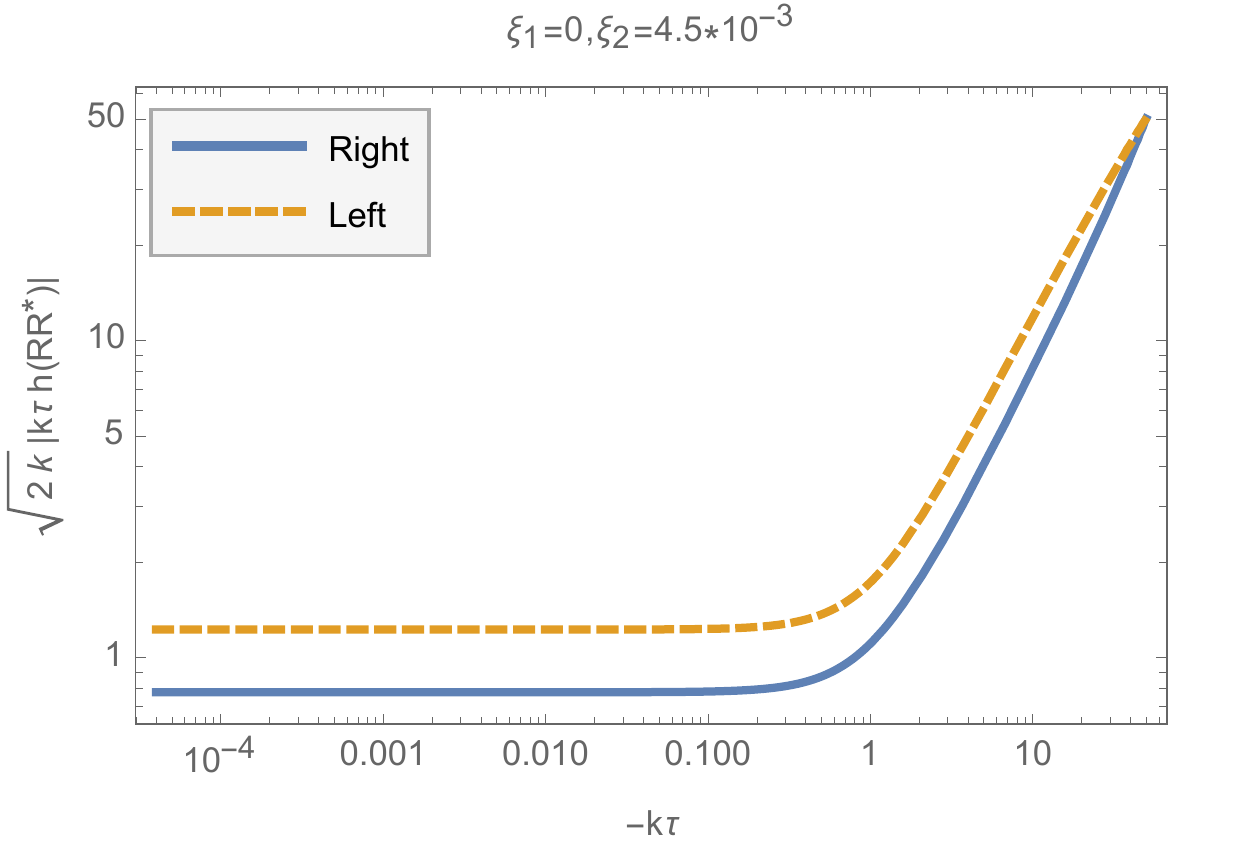} 
\end{subfigure}
\begin{subfigure}{0.5\textwidth}
\includegraphics[width=0.99\linewidth, height=5.4cm]{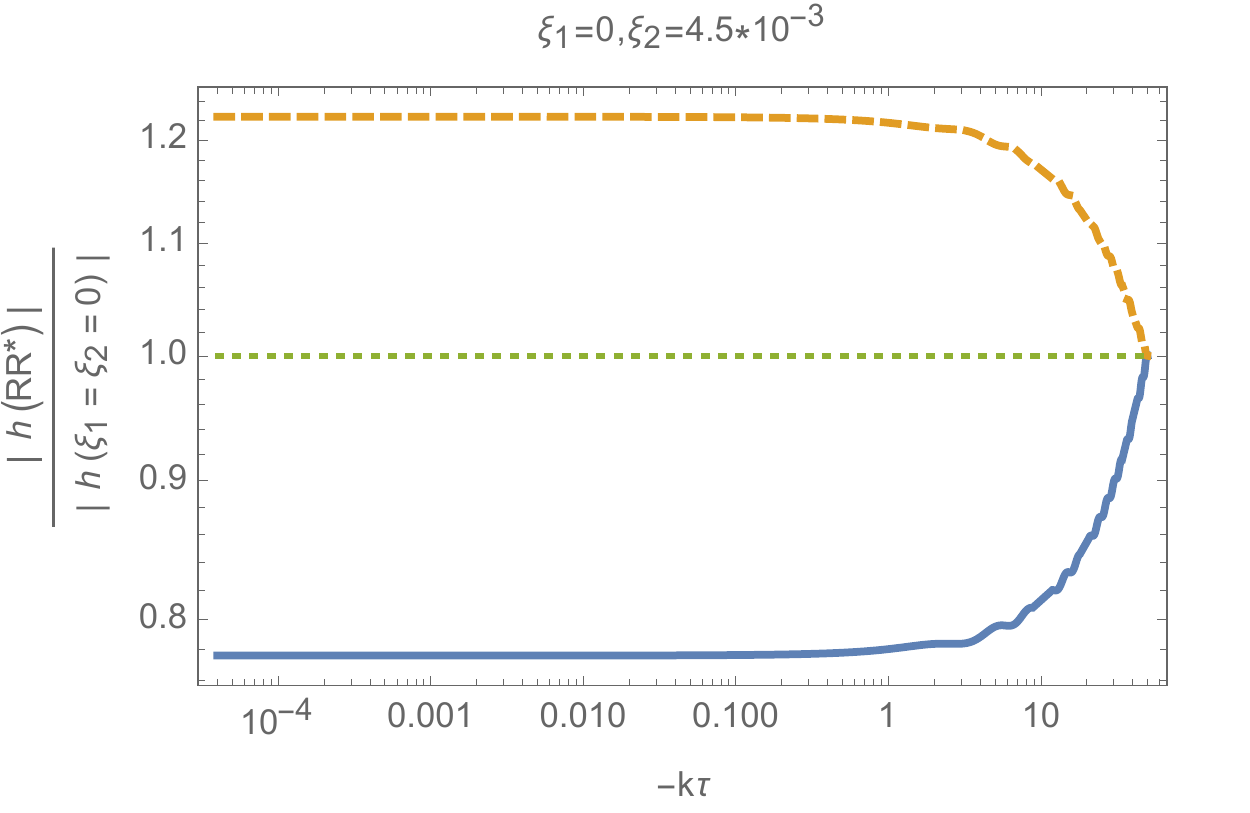}
\end{subfigure}
\begin{subfigure}{0.5\textwidth}
\includegraphics[width=0.99\linewidth, height=5.4cm]{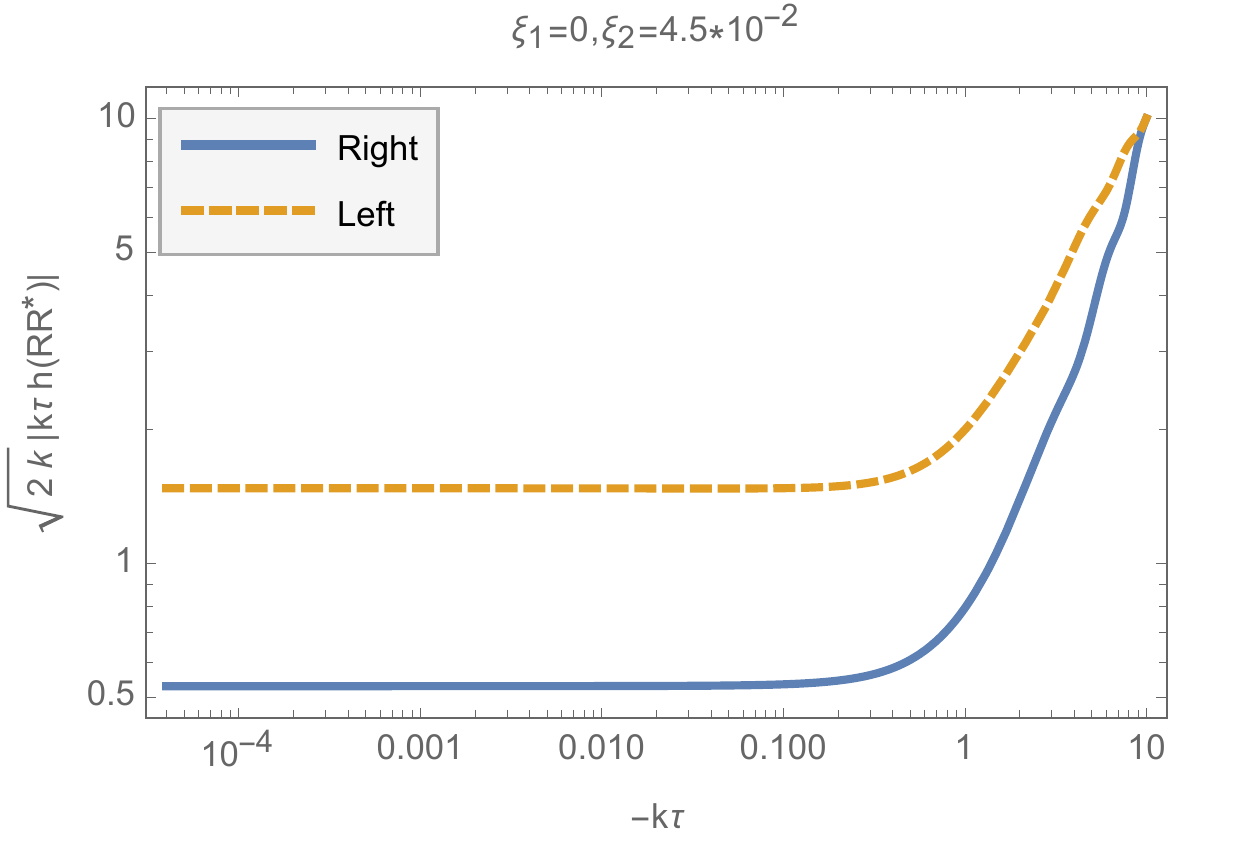} 
\end{subfigure}
\begin{subfigure}{0.5\textwidth}
\includegraphics[width=0.99\linewidth, height=5.4cm]{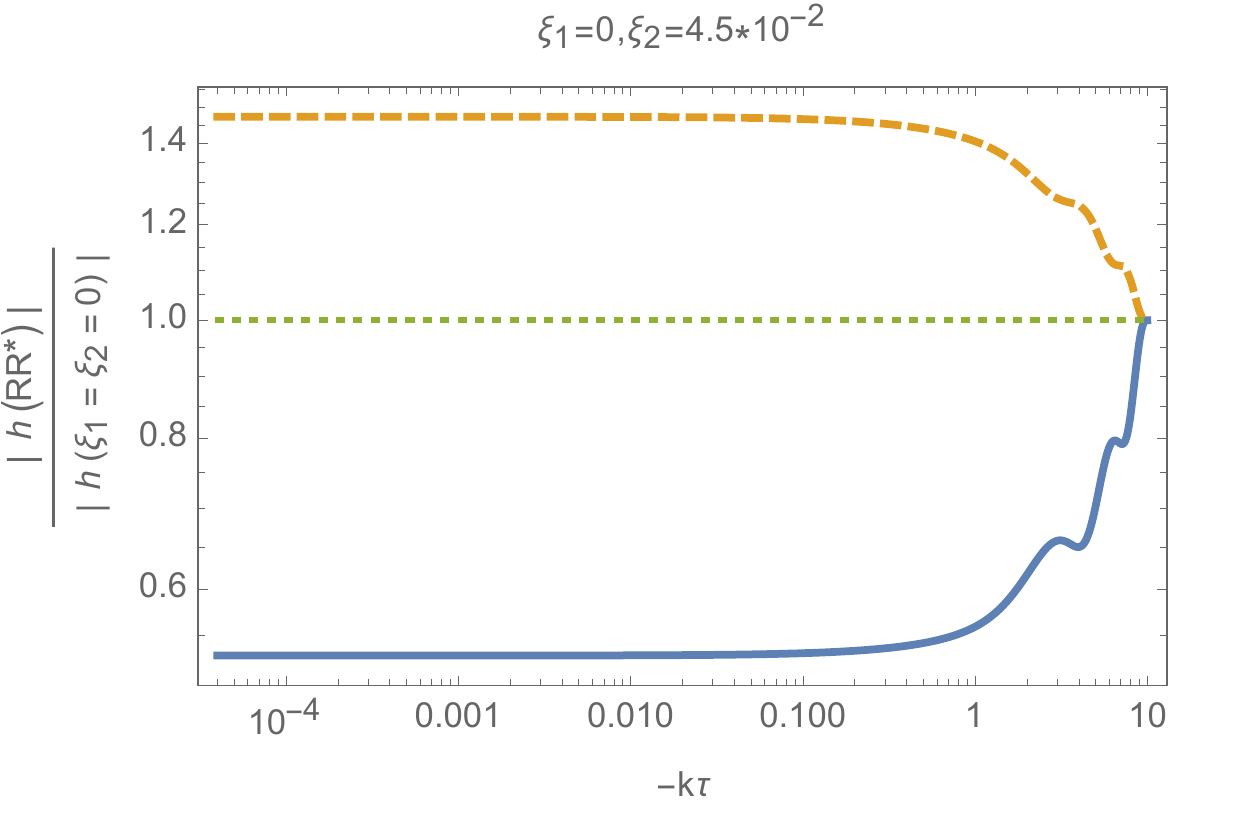}
\end{subfigure}
\caption{\textbf{(Left panel)} The right- (solid blue) and left-handed (dashed orange) helicity mode functions of $h_{R,L}$ for different values of $\xi_2$ and $\xi_1 = 0$. We plot $\sqrt{2k}|k\tau h_{R,L}|$. \textbf{(Right panel)}  The ratios of the right- (solid blue) and left-handed (dashed orange) helicity mode functions for the same values of $\xi_2$ as the left panels with respect to those for $\xi_2 =\xi_1 = 0$. The horizontal dotted line shows unity.}
\label{fig:iii}
\end{figure}

To understand the effect of $R\tilde{R}$, we first consider the case where $\xi_1= 0$ and $m_Q = 0$ in \eqref{eomhR}. The last term on the left hand side of \eqref{eomhR} acts as a \textit{friction} term for $h_L$ which prevents it from decaying, whereas it acts as an \textit{anti-friction} term for $h_R$, which makes $h_R$ decay faster. We plot the metric tensor mode functions for different values of $\xi_2$ in Figure \ref{fig:iii}. The difference between the right- and left-handed helicity modes are negligible for a small value of $\xi_2$, as shown in the top-left panel of Figure \ref{fig:iii}. As $\xi_2$ becomes larger, the difference between the two helicity modes becomes more visible (middle- and bottom-left panels). 

In the right panels of Figure \ref{fig:iii}, we show the ratios of the right- and left-handed helicity mode functions with respect to those for $\xi_1= 0$, $m_Q = 0$ and no gravitational Chern-Simons term, labelled as $h_{\xi_2 =\xi_1 = 0}$. Contrary to the case where gauge fields are present, the left-handed helicity of the metric tensor perturbations is amplified. This difference depends on the relative sign between the coefficients of the parity-violating terms $F\tilde{F}$ and $R\tilde{R}$, i.e., $\lambda_1$ in \eqref{SPEC} and $\lambda_2$ in \eqref{GCS}. The effect of $R\tilde{R}$ on the enhancement/suppression is nearly symmetric as shown in the right panels of Figure \ref{fig:iii}. This enhancement/suppression occurs already deep inside the horizon. On the other hand, amplification of the right-handed helicity of the gauge field occurs near horizon crossing (see the green dotted line in the left panel of Figure \ref{fig:i}). This difference becomes important in section \ref{FF}.

\subsection{With $F\tilde{F}$}
\label{FF}
\begin{figure}[t]
\begin{subfigure}{0.5\textwidth}
\includegraphics[width=0.99\linewidth, height=5.4cm]{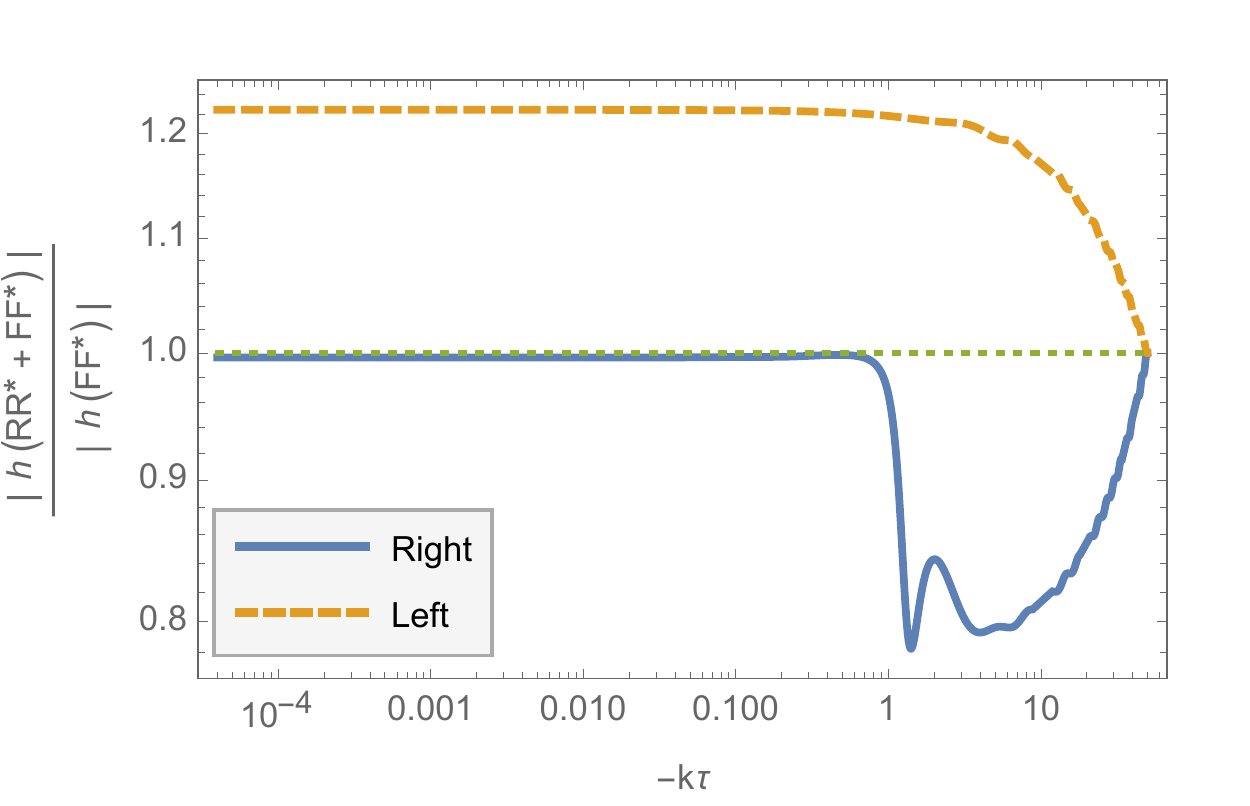} 
\end{subfigure}
\begin{subfigure}{0.5\textwidth}
\includegraphics[width=0.99\linewidth, height=5.4cm]{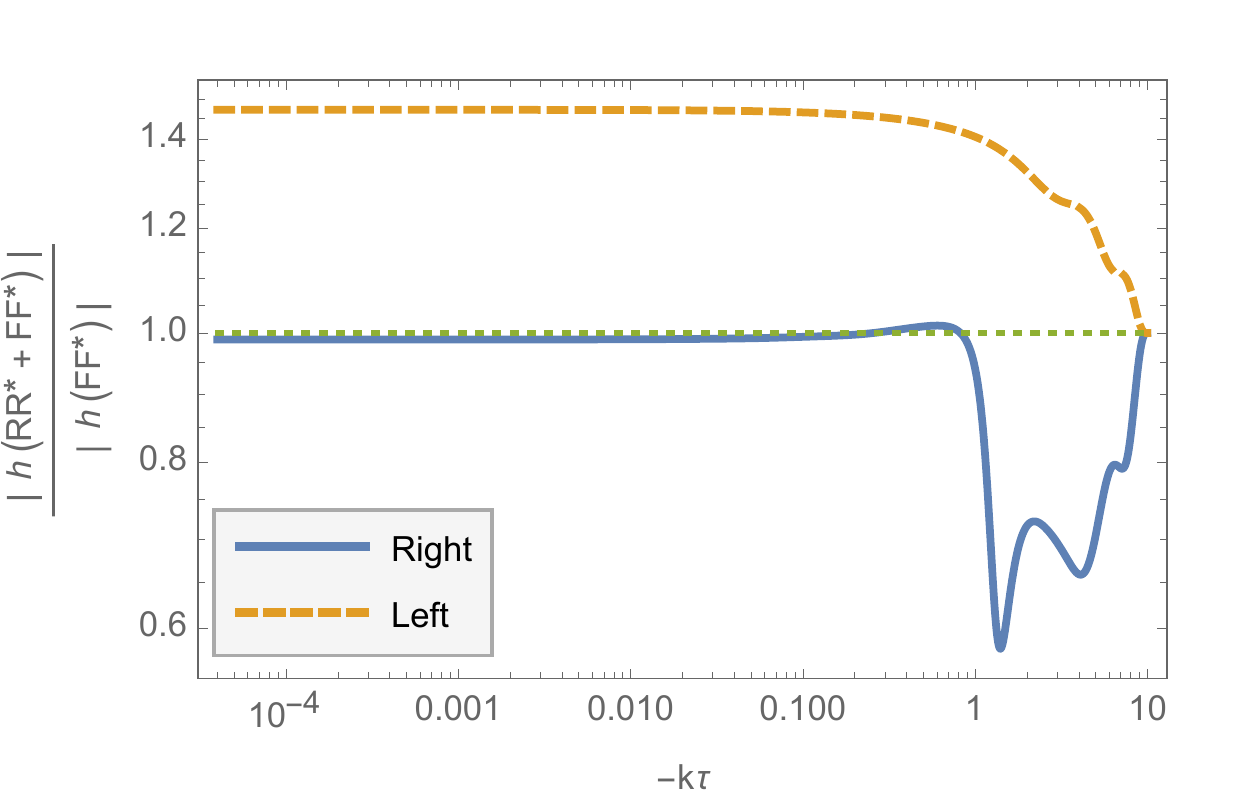}
\end{subfigure}
\caption{Same as the right panel of Figure \ref{fig:i}, but for $\xi_2 = 4.5\times10^{-3}$ \textbf{(Left panel)} and $\xi_2 = 4.5\times10^{-2}$\textbf{(Right panel)}.}
\label{fig:ii}
\end{figure}

We turn on the $F\tilde{F}$ term with $\xi_1 = 3.3$ $(m_{Q} = 3 )$. To capture the effect of $R\tilde{R}$ in axion-SU(2) gauge field models, we plot the ratio of metric tensor mode functions for different values of $\xi_2$ with respect to those without the gravitational Chern-Simons term, i.e. $\xi_2 = 0$ in Figure \ref{fig:i} and \ref{fig:ii}. In the right panel of Figure \ref{fig:i} for $\xi_2 = 4.5\times10^{-6}$, the contribution of the gravitational Chern-Simons term is small given such a small value of $\xi_2$. In Figure \ref{fig:ii} we have plotted the same as Figure \ref{fig:i} for larger values of $\xi_2$. After considering different configurations, we conclude that the contribution from the gravitational Chern-Simons term on the left-handed helicity modes is about fifty percent amplification for $\xi_2 = 4.5\times10^{-2}$ as shown in Figure \ref{fig:ii} while the right-handed helicity modes are largely unaffected. This value of $\xi_2$ requires a large hierarchy between $\lambda_2$ and $\lambda_1$, as noted at the end of section \ref{sec:bg}. 

\begin{figure}[ht!]
\begin{subfigure}{0.5\textwidth}
\includegraphics[width=0.99\linewidth, height=5.4cm]{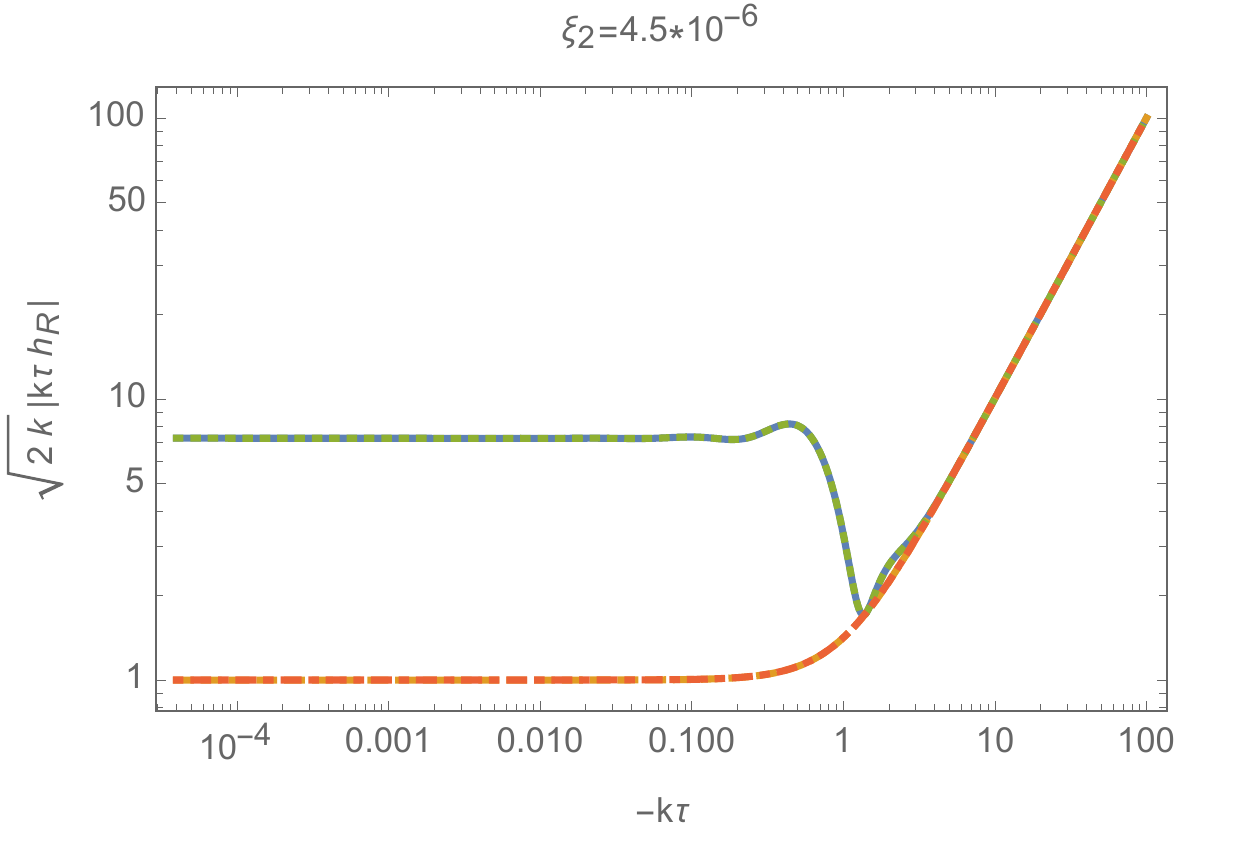} 
\end{subfigure}
\begin{subfigure}{0.5\textwidth}
\includegraphics[width=0.99\linewidth, height=5.4cm]{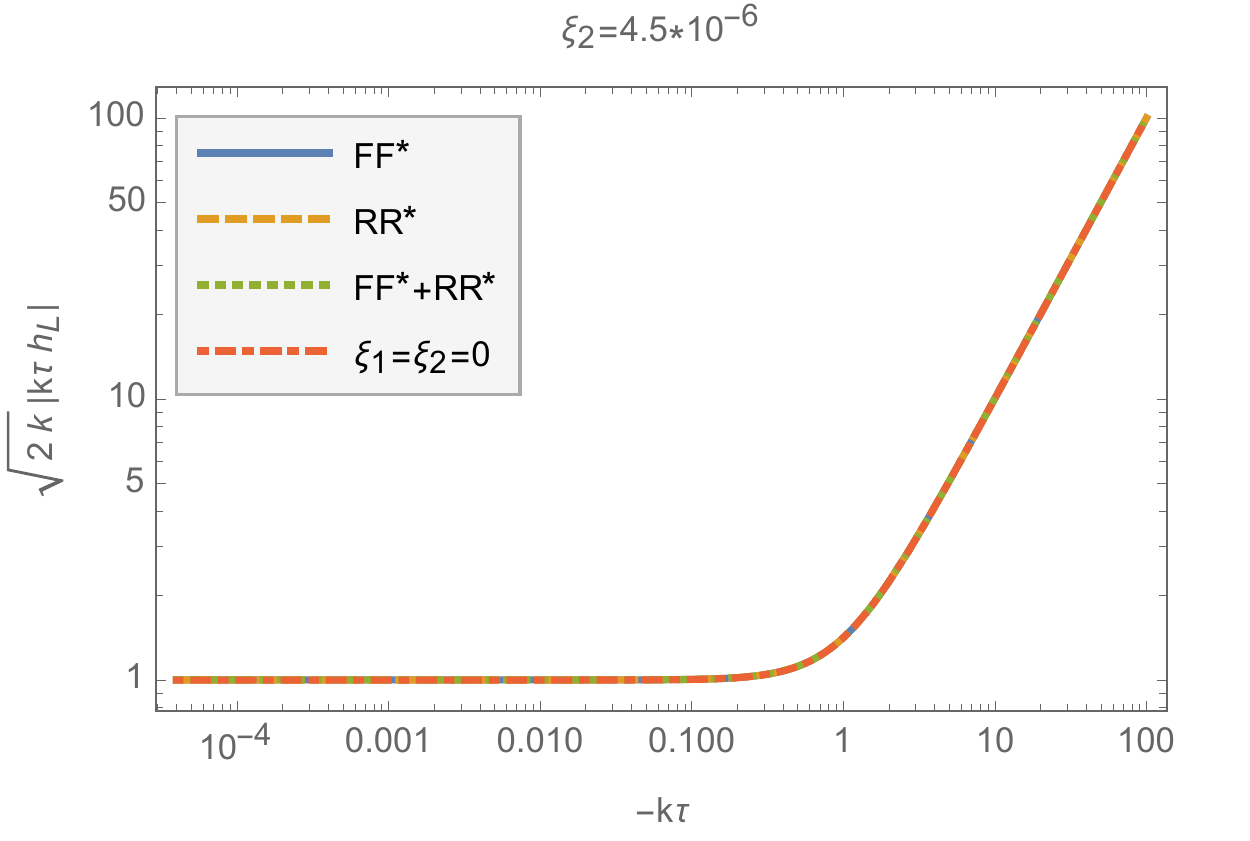}
\end{subfigure}
\begin{subfigure}{0.5\textwidth}
\includegraphics[width=0.99\linewidth, height=5.4cm]{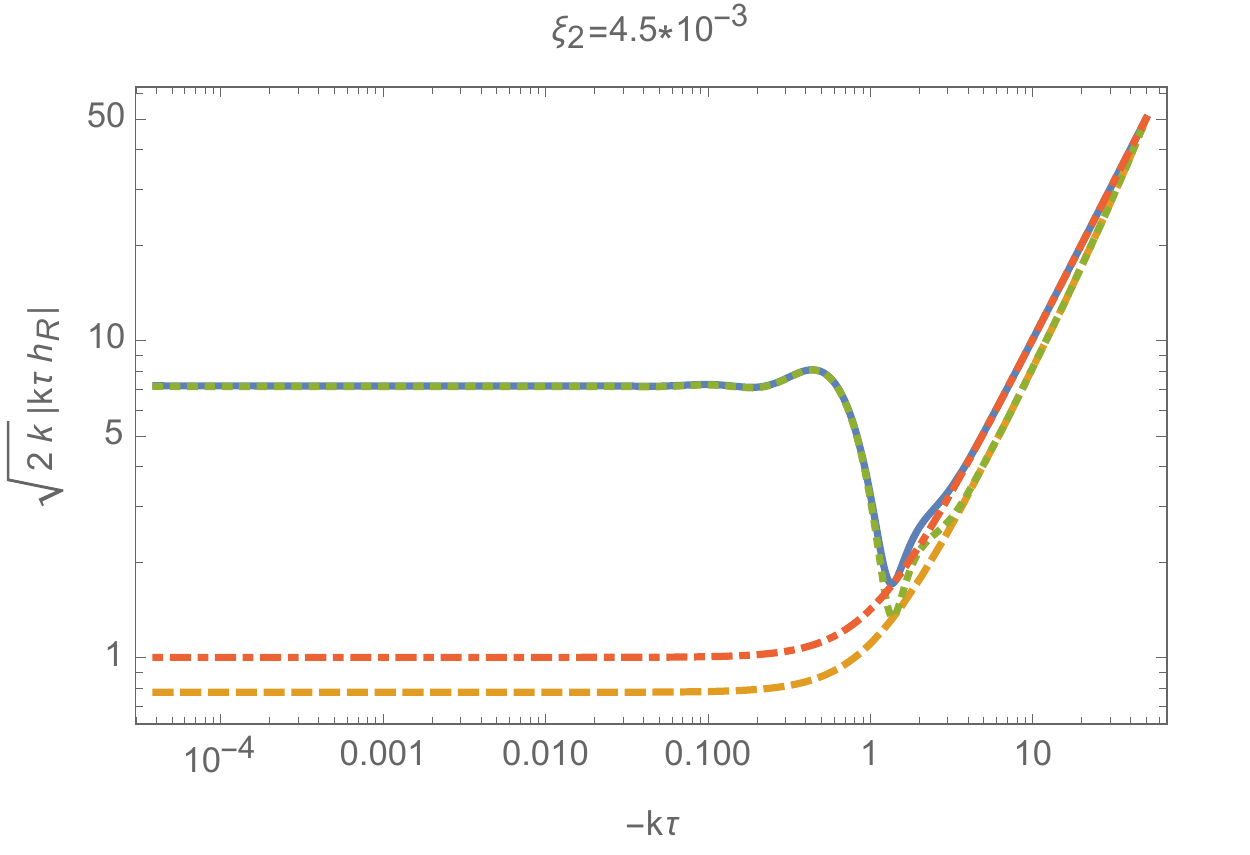} 
\end{subfigure}
\begin{subfigure}{0.5\textwidth}
\includegraphics[width=0.99\linewidth, height=5.4cm]{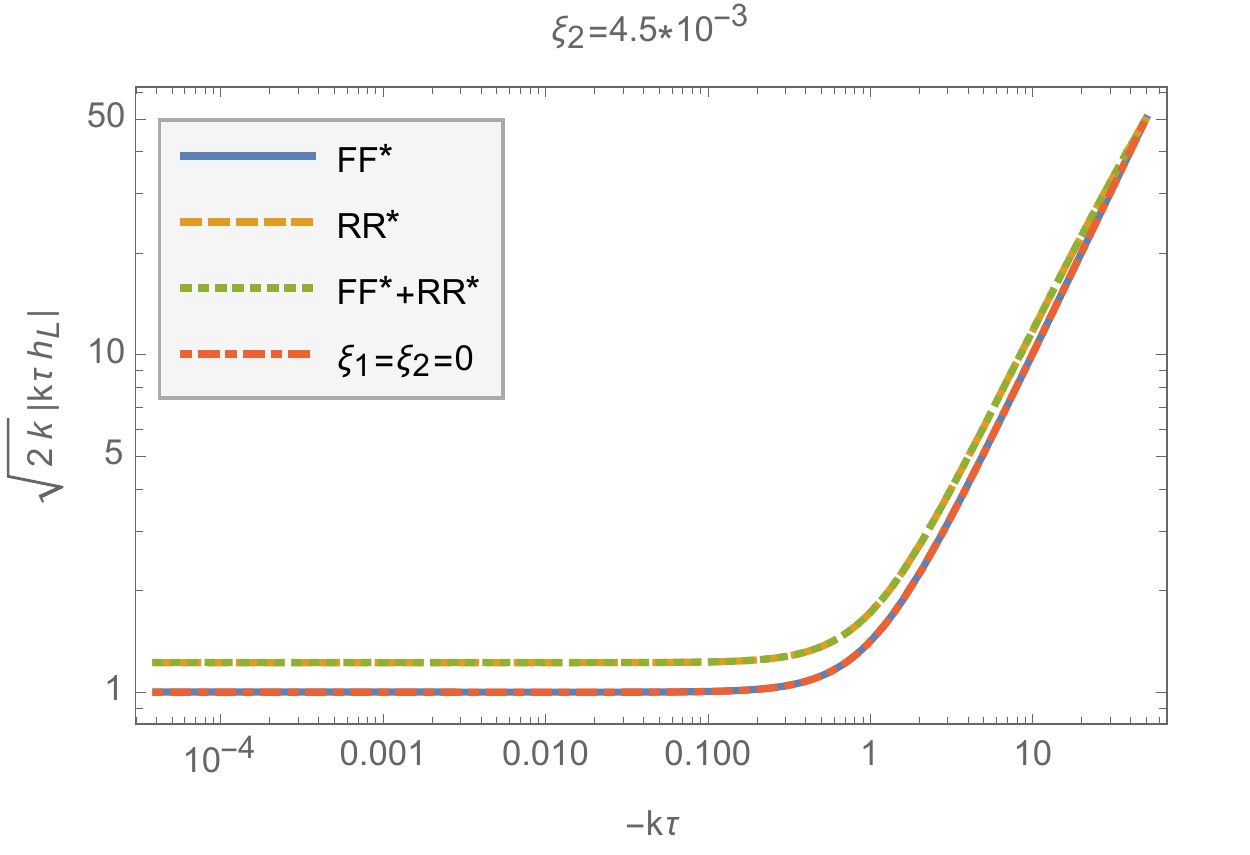}
\end{subfigure}
\begin{subfigure}{0.5\textwidth}
\includegraphics[width=0.99\linewidth, height=5.4cm]{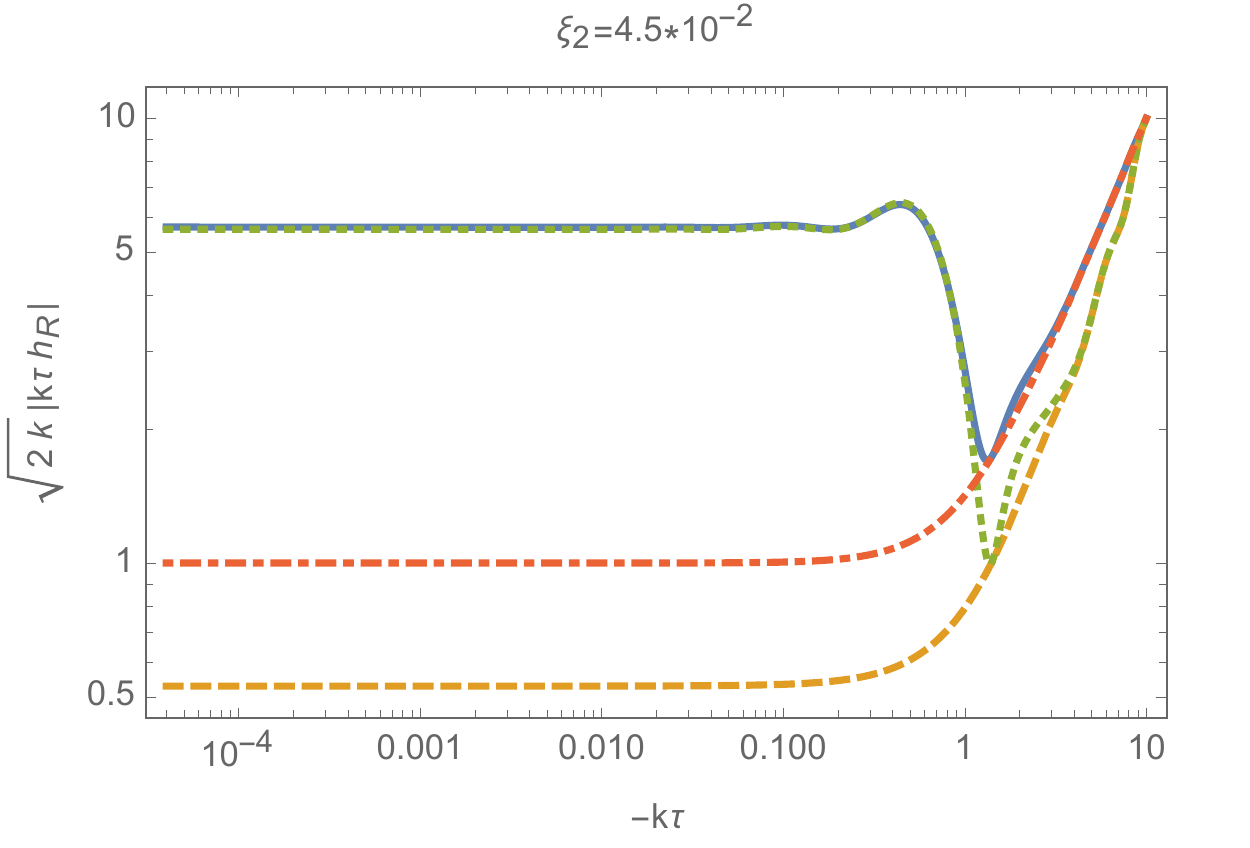} 
\end{subfigure}
\begin{subfigure}{0.5\textwidth}
\includegraphics[width=0.99\linewidth, height=5.4cm]{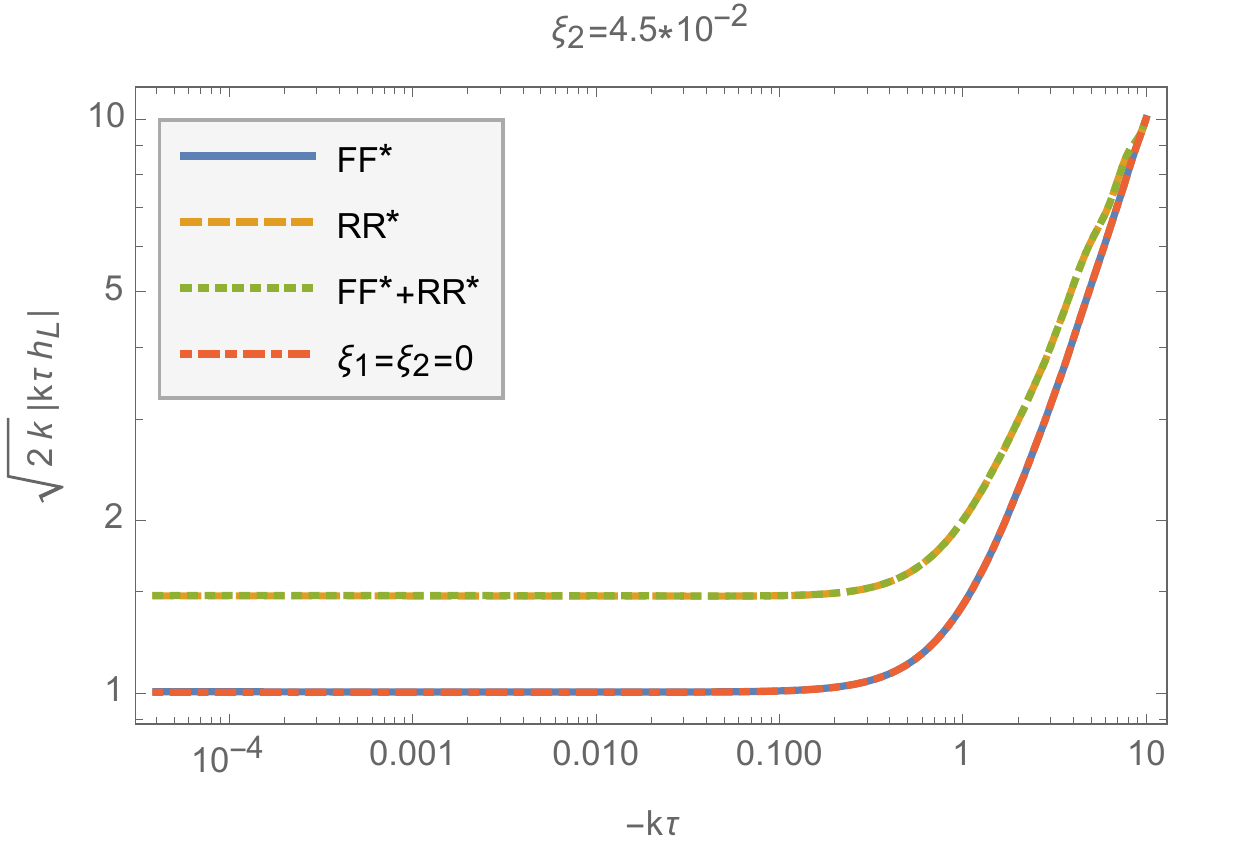}
\end{subfigure}
\caption{\textbf{(Left panel)} The right-handed helicity mode functions for four different cases: $F\tilde{F}$ and $R\tilde{R}$ (dotted green), without $R\tilde{R}$ (solid blue), without $F\tilde{F}$ (dashed orange), and without both terms (dot-dashed red) for different values of $\xi_2$. $\xi_1$ is always $\xi_1 = 3.3$ $(m_{Q} = 3)$. \textbf{(Right panel)} Same as the left panels but for the left-handed modes.} 
\label{fig:iv}
\end{figure}

For completeness, the right- and left-handed helicity mode functions for four different cases: with $F\tilde{F}$ and $R\tilde{R}$, without $R\tilde{R}$, without $F\tilde{F}$, and without both terms, for different values of $\xi_2$ are shown in Figure \ref{fig:iv}. 

The right-handed helicity modes are unaffected by the gravitational Chern-Simons coupling because they are sourced by the gauge field after horizon crossing, while the gravitational Chern-Simons coupling affects mode functions already deep inside the horizon.

\section{Stability Analysis} 
\label{sec:stab}


For $k > \mathcal{H}/\xi_2$, the sign of the kinetic term of $h_R$ in the equation \eqref{secondorderlag} becomes negative and, consequently, ghost instabilities may, in principle, be introduced into the model \cite{Alexander:2009tp, Dyda:2012rj}. Existence of ghosts does not necessarily translate to catastrophe in a model but translates to the breaking of the effective theory. 
Let us rewrite the first term in \eqref{secondorderlag}, $(1-  k \xi_2 / \mathcal{H})$ (this is the only problematic term we have to analyse), in physical coordinates. It is given by $( 1 - \frac{\xi_2}{H} k_{phy})$, where $k_{phy} \equiv k/a$ is the physical wave number. To show that the gravitational Chern-Simons term in this model is ghost-free, i.e., stable, we have to show that the effective field theory cut-off, $\Lambda$, on the physical wave number, $k_{phy}$, is below $H/\xi_2$.  Note that we have two new free parameters in this model, the gravitational Chern-Simons coefficient $\lambda_2$ in \eqref{GCS} and the cut-off $\Lambda$. As there are no \textit{a priori }constraints on $\lambda_2$, our strategy is to work our way backwards. Specifically, relying on independently motivated constraints on $\xi_1$, we ask what constraint is imposed on $\lambda_2$ in order to guarantee that the theory is healthy. Once this question is answered, we will ask how stringent or natural the resulting constraint is.

Let us first take a look at \eqref{ndparameters} and write the relation for $\lambda_2$
\Beq
\label{lambda2}
\lambda_2 = 2 \xi_2  \frac{\lambda_1}{\xi_1}  \bigg(\frac{M_{pl}}{H}\bigg)^2\,.
\Eeq
To remain in the ghost-free regime we need the cut-off $\Lambda$ on $k_{phy}$ to be the following:
\Beq
\label{boundxi2}
\frac{\xi_2}{H} \Lambda <1\,. 
\Eeq

Here we consider two cases, a conservative case where the cut-off $\Lambda$ does not exceed $M_{pl}$, and a more radical case where it is around $20 H$.\footnote{For more details on the cut-off, see Appendix \ref{initial}} \\
\begin{itemize}
\item \textbf{Conservative case:} The inequality in \eqref{boundxi2} boils down to $\xi_2 < H/M_{pl} $ \cite{Alexander:2004wk,Kamada:2019ewe}, given the assumption that $\Lambda$ does not exceed $M_{pl}$. Using this in \eqref{lambda2}, we have:
\Beq
\label{case1}
\lambda_2 < 2 \bigg(\frac{\lambda_1}{\xi_1}\bigg)\bigg(\frac{M_{pl}}{H}\bigg)\,.
\Eeq

\item \textbf{More radical case:} The inequality in \eqref{boundxi2} boils down to $\xi_2 < 1/20$, given the assumption that $\Lambda$ is around $20 H$. Using this in \eqref{lambda2}, we have:
\Beq
\label{case2}
\lambda_2 < \bigg(\frac{1}{10}\bigg) \bigg(\frac{\lambda_1}{\xi_1}\bigg) \bigg(\frac{M_{pl}}{H}\bigg)^2\,.
\Eeq
\end{itemize}

On the right hand side of both inequalities above we have $\xi_1 \sim O(1)$, which guarantees a slow variation of the gauge field, $\lambda_1 \sim O(10)$, and there is an upper bound on the tensor-to-scalar ratio $r \equiv (P_h / P_{\zeta}) <0.06$ from not observing tensor modes in the CMB \cite{Ade:2018gkx} where $P_h$ and $P_{\zeta}$ are the power spectra of tensor and curvature perturbations, respectively. In our model both the vacuum fluctuations of the metric and the sourced gravitational waves contribute to $P_h$. Using the upper bound on $r$, the measurement of the dimensionless power spectrum of scalar fluctuations, $\Delta_{\zeta} \equiv k^3  P_{\zeta} / 2\pi^2 \approx 2.2 \times 10^{-9}$, and the expression for the dimensionless power spectrum of tensor fluctuations only from the metric vacuum fluctuations $\Delta_{h vac} \equiv k^3  P_{h vac} / 2\pi^2 = 2 H^2 / (\pi^2 M_{pl}^2)$, we get a bound on the last term $(M_{pl}/H)^2 \gtrsim 1.5 \times 10^{9}$. 

Therefore, in both \eqref{case1} and \eqref{case2}, the right side is expected to be a very large number. As there is no stringent constraint on the free parameter $\lambda_2$ in our model, the model is not disfavoured by fine-tuning arguments.

\section{Discussion}
\label{sec:dis}
We have studied the effect of the gravitational Chern-Simons term coupled to the axion field on production and propagation of gravitational waves during inflation with the spectator axion-SU(2) sector \cite{Dimastrogiovanni:2016fuu}. Both parity-violating terms $R\tilde{R}$ and $F\tilde{F}$ exist simultaneously. 

We find that the effect of the $R\tilde{R}$ term on chiral gravitational waves can be as large as fifty percent amplification for the left-handed helicity mode functions compared to the case without the $R\tilde{R}$ term for $\xi_2 = 4.5 \times 10^{-2}$. The effect is smaller for smaller values of $\xi_2$. The right-handed helicity mode functions are unaffected regardless of the values of $\xi_2$. Moreover, using the existing bounds on $m_Q$ and $\xi_1$ from the spectator axion-SU(2) gauge field sector, and requiring that the cut-off scale of the theory, $\Lambda$, is in the conservative case $\Lambda = M_{pl}$ and in a more radical case $\Lambda = 20 H$, we put constraints on the new free parameter $\lambda_2$ in our model to remain in the ghost-free regime. Consequently, values of $\xi_2$ are related to the cut-off scale of the theory, $\Lambda$. $\xi_2 = 4.5 \times 10^{-2}$ is allowed when $\Lambda = 20 H$ and $\xi_2 = 4.5 \times 10^{-6}$ is allowed when $\Lambda = M_{pl}$.

We conclude that the inflation models with the spectator axion-SU(2) sector remain phenomenologically viable in the presence of the gravitational Chern-Simons term.

\acknowledgments
LM thanks Valerie Domcke for useful discussions and is grateful to Elisa Ferreira and Ryo Namba for insightful discussions and comments on the manuscript. EK thanks Azadeh Maleknejad for useful discussions. This reasearch was supported in part by the Excellence Cluster ORIGINS which is funded by the Deutsche Forschungsgemeinschaft (DFG, German Research Foundation) under Germany's Excellence Strategy -EXC-2094-390783311. YW is supported by JSPS KAKENHI Grant No. JP16K17712.

\appendix 
\section{Quantisation of the tensor modes and initial conditions}\label{initial}

Since the tensor-like perturbations in the gauge fields and the tensor metric perturbations are linearly coupled, we expand both in terms of the same pair of creation and annihilation operators \cite{Weinberg:2008zzc}
\Beq
h_A(\tau, k)&=\sum_{n=h,t}\left[a^A_{n,k}h_{A,n}+a^{A\dagger}_{n,-k}h_{A,n}^*\right]\,,\\
t_A(\tau, k)&=\sum_{n=h,t}\left[a^A_{n,k}t_{A,n}+a^{A\dagger}_{n,-k}t_{A,n}^*\right]\,,
\Eeq
where we have the standard commutators
\Beq
\left[a^A_{n,k},a^{B\dagger}_{m,q}\right]=\delta_{n,m}\delta_{A,B}\delta^{(3)}(k-q)\,.
\Eeq
We have to specify the vacuum (when $-k\tau\gg m_Q$ and $-k\tau\xi_2<1$) for the gravitational and gauge fields separately. We divide the vacuum space to two subspaces corresponding to each field as \cite{Dimastrogiovanni:2012ew,Namba:2013kia, Caldwell:2017chz}: 
\Beq
h_{A,n}(\tau, k) =\frac{1}{\sqrt{2k}}\delta_{n,h}e^{-ik\tau}\,,\quad h'_{A,n}(\tau, k)= -\frac{1}{\sqrt{2k}}ik \delta_{n,h}e^{-ik\tau}\,,\\
t_{A,n}(\tau, k) =\frac{1}{\sqrt{2k}}\delta_{n,t}e^{-ik\tau}\,, \quad t'_{A,n}(\tau, k)= -\frac{1}{\sqrt{2k}}ik \delta_{n,t}e^{-ik\tau}\,. 
\Eeq
The solution $h_{A,n=h}$ can be interpreted as the vacuum gravitational wave, whereas $h_{A,n=t}$ as the sourced one (by the vacuum gauge field, $t_{A,n=t}$).

While choosing a cut-off for the gravitational Chern-Simons term we should note where the tachyonic instability in the gauge sector exists for a given $m_Q$. Considering $m_Q = 3$ in the equation of motion for the gauge fluctuation, the tachyonic instability takes place around $x \sim 10$. A reasonable cut-off must be chosen far enough to capture the effects of the instability completely. The cut-off $\Lambda = 20H$ is acceptable considering this criteria.

\bibliographystyle{JHEP}
\bibliography{mybiib}

\end{document}